\magnification = \magstep1
\pageno=0
\hsize=15.truecm
\hoffset=0.5truecm
\vsize=22.0truecm

\output={\plainoutput}
\pretolerance=3000
\tolerance=5000
\hyphenpenalty=10000    
\newdimen\digitwidth
\setbox0=\hbox{\rm0}
\digitwidth=\wd0

\def\footnoterule{\kern-3pt \hrule width \hsize \kern 2.6pt
\vskip 3pt}
\def\cl{\centerline}
\def\ni{\noindent}

\def\msun{M$_{\odot}$}

\def\vs{\vskip 11pt}

\def\solar{\ifmmode _{\mathord\odot}\else $_{\mathord\odot}$\fi}

\font\ksub=cmsy7
\def\teff{T$_{\kern-0.8pt{\ksub e\kern-1.5pt f\kern-2.8pt f}}$}
%
\vs\vs\vs
\vs\vs\vs
\vs
\vs
\vs
\cl{\bf AN OUTLINE OF RADIATIVELY-DRIVEN COSMOLOGY}
\vs\vs
\cl{Robert L. Kurucz}
\cl{Harvard-Smithsonian Center for Astrophysics}
\vs
\cl{December 13, 1991}
\cl{Revised January 24, 1993}
\cl{Revised October 2, 1993}
\cl{Revised October 19, 1994}
\cl{Revised May 8, 1997}
\cl{Revised March 5, 2000}
\eject
\vs\vs\vs
\cl{\bf AN OUTLINE OF RADIATIVELY-DRIVEN COSMOLOGY}
\vs\vs
\cl{Robert L. Kurucz}
\vs
\cl{Harvard-Smithsonian Center for Astrophysics, 60 Garden St, Cambridge, MA 02138}
\vs\vs\vs
 
\cl{ABSTRACT}\vs

A Big Bang universe consisting, before recombination, of H, D, $^{3}$He, 
$^{4}$He, $^{6}$Li, and $^{7}$Li ions, electrons, photons, and massless 
neutrinos, at closure density, with a galaxy-size perturbation spectrum but 
no large-scale structure, will evolve into the universe as we now observe it.  
Evolution during the first billion years is controlled by radiation.  
Globular clusters are formed by radiatively-driven implosions, galaxies are 
formed by radiatively triggered gravitational collapse of systems of globular 
clusters, and voids are formed by radiatively-driven expansion.  After this 
period the strong radiation sources are exhausted and the universe has 
expanded to the point where further evolution is determined by gravity and 
universal expansion.
 
\ni Subject headings: cosmology --- stars: Population III --- stars: Population II ---
clusters: globular --- galaxies: evolution
\vs
\vs
\cl{ 1. INTRODUCTION}
\vs
     Cosmology suffers from the same sort of conceptual error as did geology
and evolutionary biology earlier in this century. ``Gradualism" or
``uniformitarianism" and slow changes are assumed, probably because it makes
modeling easier.  ``There was the Big Bang.  There was decoupling
(= recombination).  Nothing much else has happened.  Gravity is the only
force that matters.  Evolution is proceeding slowly and only a 
fraction of matter has formed galaxies and stars.  The ``microwave" background
just sits there.  The only important science is determining the expansion
parameters".  Gradual evolution has always turned out to be a delusion
produced by oversimplification.
 
     In reality, a second force is produced by radiative acceleration.
It triggers rapid collapses that go to almost 100\% completion.  It produces
``catastrophic" or ``episodic" evolution.  
 
     Here we present the results of gedanken experiments
(Kurucz 1992) in a traditional, linear, chronological sequence in the hope
of stimulating research on the many topics considered.
\vfill
\eject
\cl{ 2. CONDITIONS BEFORE RECOMBINATION}
\vs
     The evolution of the universe from before recombination
to the present time can be explained by simple, elementary physics.
Let us start when the universe is a few hundred thousand years old, at the 
time when the temperature has fallen to about 10000K.  Let it consist of 
H, D, $^{3}$He, $^{4}$He, $^{6}$Li, and $^{7}$Li ions, electrons, photons, 
and massless neutrinos at the closure density, between 10$^{4}$ and 
10$^{5}$ per cubic centimeter.  Abundances are taken from standard Big Bang
nucleosynthesis calculations shown in Figure 1.  These abundances are
ten times higher for Li, 10 times lower for  $^{3}$He and D than cosmologists 
have assumed in the past but they are consistant with observation (He, Sasselov 
and Goldwirth 1995; D upper limit, Lubowich et al 1994; Li, Kurucz 1995) 
in that there are no observations of primordial $^{3}$He or D, and in that 
the Li abundance in extreme Population II stars has been grossly underestimated.
 
    The gas is opaque.  The redshift Z is approximately 1300.  There is
uniform expansion and cooling of the universe.  There is no large-scale
structure; the universe is filled with galaxy-size perturbations in density 
and temperature that were created at an earlier time.  As the universe 
expands those perturbations evolve into highly structured galaxies with 
myriad condensations, and the galaxies themselves form large-scale structures.
\vs
\cl{3. GALAXY-SIZE PERTURBATIONS}
\vs
Ignoring mergers and collisional destruction, every galaxy extant
corresponds to a prerecombination perturbation, and vice versa.  Thus the
distribution
function for the perturbation masses is approximately the distribution function 
for galaxy masses now, except at the extremes.  There are no symmetries in the
initial galaxy-size perturbations.  They have facets, convexities, concavities,
etc. from early close packing (as in a Voronoi tesselation).  Because
there is no symmetry, every perturbation has angular momentum.

     The perturbations are in quasi-hydrostatic equilibrium with gravity
pulling inward trying to increase the density while radiative acceleration
pushes outward trying to smooth out the perturbation.  The cosmological 
expansion enhances the perturbation.  The denser, hotter regions are compressed 
(i.e., they expand less rapidly) while less dense regions are pulled apart.  
The local gravity vector g does not point radially toward the perturbation 
``center".  The radiative acceleration vector g$_{\rm rad}$ has similar 
components pointing in the  opposite direction.  The effective gravity at 
any point is g$_{\rm eff}$ = g + g$_{\rm rad}$ .  The surface and volume of 
the perturbations are defined by the surfaces  g$_{\rm eff}$ = 0.  
 
     From this starting point the universe continues to expand and cool until
the temperature drops to a few thousand degrees.  The electrons combine with
the ions until most of the matter is neutral or negative.  The opacity of the 
gas drops and radiative acceleration plummets.
\vfill
\eject
\cl{ 4. FORMATION OF ATOMS AND MOLECULES}
\vs
 Recombination actually starts as soon as electrons and protons are formed.
What happens at ``recombination" or ``decoupling" is that photoionization
(of hydrogen) stops.  The electron number density drops drastically so that
the gas pressure drops by a factor of 2.  The electron contribution to the 
opacity drops drastically as does the radiative acceleration. 
 
     Recombination and cooling is much more complicated than has been assumed.
The recombinations in order of energy are
 
$^{7}$Li$^{+++}$ + e $>$ $^{7}$Li$^{++}$ +   987660 cm-1
 
$^{6}$Li$^{+++}$ + e $>$ $^{6}$Li$^{++}$ +   987647 cm-1
 
$^{7}$Li$^{++}$  + e $>$ $^{7}$Li$^{+}$  +   610080
 
$^{6}$Li$^{++}$  + e $>$ $^{6}$Li$^{+}$  +   610066
 
$^{4}$He$^{++}$  + e $>$ $^{4}$He$^{+}$  +   438909
 
$^{3}$He$^{++}$  + e $>$ $^{3}$He$^{+}$  +   438889
 
$^{4}$He$^{+}$   + e $>$ $^{4}$He   +   198311
 
$^{3}$He$^{+}$   + e $>$ $^{3}$He   +   198291 ?
 
$^{2}$H$^{+}$    + e $>$ $^{2}$H    +   109709
 
$^{1}$H$^{+}$    + e $>$ $^{1}$H    +   109679
 
$^{7}$Li$^{+}$   + e $>$ $^{7}$Li   +    43487
 
$^{6}$Li$^{+}$   + e $>$ $^{6}$Li   +    43472
 
$^{2}$H   + e $>$ $^{2}$H$^{-}$    +   6061+ ?
 
$^{1}$H   + e $>$ $^{1}$H$^{-}$    +   6061
 
$^{7}$Li   + e $>$ $^{7}$Li$^{-}$    +   4981   
 
$^{6}$Li   + e $>$ $^{6}$Li$^{-}$    +   4981- ? 
 
At the same time, there may also be high temperature molecules: all the positive 
and negative ions of Li$_{2}$, LiHe, LiH, He$_{2}$, HeH, and H$_{2}$ and their 
isotopomers.  For example,
 
$^{7}$Li$^{1}$H$^{++}$ $^{7}$Li$^{1}$H$^{+}$ $^{7}$Li$^{1}$H $^{7}$Li$^{1}$H$^{-}$

$^{7}$Li$^{2}$H$^{++}$ $^{7}$Li$^{2}$H$^{+}$ $^{7}$Li$^{2}$H $^{7}$Li$^{2}$H$^{-}$

$^{6}$Li$^{1}$H$^{++}$ $^{6}$Li$^{1}$H$^{+}$ $^{6}$Li$^{1}$H $^{6}$Li$^{1}$H$^{-}$

$^{6}$Li$^{2}$H$^{++}$ $^{6}$Li$^{2}$H$^{+}$ $^{6}$Li$^{2}$H $^{6}$Li$^{2}$H$^{-}$

Passing through each He recombination reduces the number of particles and the 
gas pressure by 5\%.  The H recombination reduces the number of particles and 
the gas pressure by 45\%.
Li remains partially ionized and provides free electrons which can form
H$^{-}$ and Li$^{-}$.  It can also participate in charge exchange reactions.  

Decoupling is never complete because there are free electrons from the Li that
Thomson scatter, because H and He Rayleigh scatter, because Li has lines in the
visible that are optically thick on globular cluster scales, and because H$^{-}$
has continuous absorption in the visible and infrared that is optically thick
at galaxy scales.  The universe is optically thick to the recombination radiation.
Thus the ``microwave" background is not from the primordial black body but from
a later time.
\vfill
\eject
\cl{ 5. FORMATION OF GLOBULAR-CLUSTER-SIZE PERTURBATIONS}
\vs
   When the radiation field suddenly decouples, g$_{\rm rad}$ becomes small
and P$_{\rm gas}$ collapses by a factor of more than 2,
and g$_{\rm eff}$ suddenly, impulsively increases to g.  This inward impulse
produces waves that travel at the speed of sound.
However, because there is no symmetry, these waves cannot behave coherently.
They cannot propagate far before interacting with other waves.  They interfere
in three dimensions.  Perhaps they form shocks.  The
globular-cluster-size perturbation spectrum that they produce has high-density,
low-mass maxima and low-density minima, all superimposed on the galaxy-size
perturbation (Figure 2).   At this stage every point in the universe has two peculiar 
velocity components: one toward the local globular-cluster-size perturbation 
maximum and one toward the local galaxy-size perturbation maximum.  Research is
needed to find out whether the waves leave behind microturbulent motions in 
the perturbations.

     The temperature changes in the new perturbations are spectacular.
In the less dense regions the temperature drops.  In the dense centers
the gas heats and partially ionizes.  The opacity increases.  Positively
and negatively charged atoms and molecules flourish and radiate through the
cool surface.  As soon as the ``recombination" or ``decoupling" era begins
it is over.  The background blackbody radiation is completely destroyed.
The radiation field comes from globular-cluster-size perturbations irradiating 
each other.

 The universal expansion amplifies perturbations.  Minima become relatively
wider and maxima become sharper, both on the galaxy-size scale and on the
globular-cluster-size scale.  The universal expansion naturally separates the 
galaxy-size-perturbations and produces surfaces through which there can be 
outward flux.  This also happens with the globular-cluster-size perturbations, 
and the outermost globular-cluster-size perturbations can radiate out of the
galaxy-size perturbations and thus cool more rapidly than interior perturbations.
 
  Coldness is a modern invention.  The temperature of any matter never
got below 500K, say, until the initial Population II stars produced dust
by mass loss.  The physics of the contemporary interstellar medium is not
relevant at early times.

\vfill
\eject
\cl{ 6. FORMATION OF POPULATION III STARS}
\vs
      The universe expands by a factor of 100 from recombination, say z = 1300,
to Population III star formation, say z = 13 .  The background radiation 
produced by the collapsing perturbations cools proportionally and fills the 
expanded volume.  This radiation is always coupled to the perturbations.  Even
when it is redshifted by a factor of 100, it is still absorbed by molecules 
in the perturbations.

     Li and any heteronuclear molecules have lines in the visible and infrared.
There are between 300,000 and 400,000 lines: electronic, vibrational-rotational, 
and rotational.  The red-shifted background radiation produces an overpopulation
of the excited levels.  The excited levels can absorb radiation and then
emit at higher frequencies that are not likely to be absorbed by the 
cooler surface.  This mechanism allows the perturbation to get rid of 
excess energy from the collapse.  There are likely to be fluoresences that 
couple the different species and produce energy redistributions.  The line 
opacity may be enhanced by the high microturbulent velocity.  Differential 
velocities from the collapse can reduce or enhance absorption and emission.

     The perturbations range in mass from more than 100 \msun\ to 
10$^{6}$\msun\ .  The perturbations can collapse only as fast as excess 
energy can escape in radiation.  A small perturbation radiatively cools faster 
than a large perturbation because it has a larger surface to volume ratio.  
The outermost perturbations radiate mostly into open space between
the galaxy-size perturbations.  The smallest perturbations collapse to form,
say, 100 \msun\ Population III stars.  
 
\vfill
\eject
\cl{ 7. FORMATION OF GLOBULAR CLUSTERS}
\vs
 
     Massive Population III stars are superluminous.  They radiate
about 10$^{53}$ ergs in 10$^{6}$ years and then explode as supernovas.  These
are the only Population III stars and only their dead supernova remnants now
remain, amounting to only a small fraction of the mass of the universe.  Because
there is not enough time for larger perturbations to evolve, all other matter
in the universe is contaminated by the supernovas and becomes Population II
material.
 
     No matter what the perturbation spectrum, the big perturbations will in
general be surrounded by small perturbations.  These might have masses as
small as 100 \msun.  In diameter, these are only 20 times smaller than a
10${^6}$ \msun\  perturbation and 50 times smaller than a 10$^{7}$ \msun\
perturbation.  The radiative acceleration from each Population III star
contributes to the radiatively-driven implosion of all its neighboring
perturbations into globular clusters.  Four Population III stars tetrahedrally
arranged may be sufficient to implode the largest perturbations.
 
     Globular cluster formation happens in layers like an onion.  
The surface of a perturbation is compressed and
contaminated by the Population III stars.  It becomes optically thick and
forms a layer of Population II stars and becomes optically thin again.  Simple
versions of this process for radiatively imploding bumps on the surface of a
molecular cloud and for radiatively imploding a small cloud between two hot stars
have been presented in a series of papers by Sandford, Whitaker, and Klein
(Sandford, Whitaker, and Klein 1982; 1984; Klein, Sandford, and Whitaker 1983), 
Figure 3, but they never extrapolated the idea to the formation of a globular cluster.
Any leftover material in the outer shell is driven inward.  The layering process
repeats inward until all the matter in a large perturbation is formed into stars.
The stellar abundances and masses are determined by the number and proximity of
the supernovas.  The distribution function of these Population II masses is the
initial mass function.  The masses can range over the whole spectrum but
because the Population II material has higher opacity than the Population III
material, and because its collapse is helped along by external forces, the
masses are smaller than the Population III masses and can even be quite small.
However, the smallest Population II stars are still larger than the smallest
(future) Population I stars which form easily because of high opacity gas and
dust.  There are no initial Population II brown dwarfs.
 
     A globular cluster can be formed at any time in any population.  The only
requirement is the existence of hot stars surrounding and radiatively imploding
a large cloud.
\vfill
\eject
\cl{ 8. FORMATION OF GALAXIES}
\vs
 
     Asymmetries in the distribution of the Population III stars around each
large perturbation produce a small, net globular cluster velocity.  Since
there are excess Population III stars at the surface of galaxy-size 
perturbations, the globular clusters near those boundaries will be accelerated 
away from the boundaries and will have velocities inward on the order of a 
fraction of a km s$^{-1}$.  This is the radiative trigger that leads to the 
gravitational implosion (violent relaxation) of the systems of globular 
clusters into elliptical galaxies.  Figure 4 shows a schematic calculation 
of such violent relaxation.  As galaxy-size perturbations have no symmetry, 
they have angular momentum and they spin up as they collapse.
 
     At this point at z $\sim$ 10 we have a statistically uniform universe 
filled with elliptical galaxies.  The elliptical galaxies are transparent and 
widely spaced, but any line of sight intersects many galaxies.  
For the first time the universe becomes transparent.  The ``microwave" 
background comes either from some subsequent event in galaxy-quasar 
evolution that produces tremendous power near 100$\mu$m, or from the pair 
annihilation of background neutrinos integrated from transparency until now,
or from both.

     All of the globular clusters in these elliptical
galaxies are the same age.  The globular clusters collide and gain internal
energy and rapidly disintegrate.  By today 99.9\% of them have disintegrated.
The clusters that are left are not typical or representative of the properties
of the initial ensemble.  They were the cold tail.  They are not pure, having 
added and lost stars through their whole lives.  The current members of one of 
these globular clusters are not necessarily siblings, coeval, or even 
Population II.  There can be dark globular clusters in which all or almost 
all the stars are neutron stars and white dwarfs.  
 
     Both globular cluster formation and galaxy formation produce intergalactic
Population II gas and stars as leftovers or as high velocity ejecta.
These stars may now be main sequence dwarfs, luminous giants,
white dwarfs, or neutron stars.  Galaxy formation also produces intergalactic
globular clusters because high velocity clusters can be ejected in the violent
relaxation.
 
     Figures 5 through 12 schematically describe galactic evolution.

     If the initial mass functions of the globular clusters that form
an elliptical galaxy have almost all low mass stars, the galaxy remains
an elliptical galaxy forever.  These galaxies have low luminosity until
the giant branch is strongly populated.  A few, more massive, stars 
lose enough mass to fill the galaxy with the tenuous gas that produces 
the Lyman $\alpha$ forest.
 
     If the initial mass functions of the globular clusters have mostly
high mass stars, the elliptical galaxy evolves into a spiral galaxy.
Supernova remnants and the mass lost by intermediate mass supergiants
collapse into a bulge and a disk, which spin up.
 
     An intermediate case produces an irregular or ``young" galaxy.  

     When there is a significant high mass tail, after some 20 million years,
the whole elliptical galaxy fills with supernovas and supernova remnants.  
The galaxy fills with jumbled magnetic structures.  The galaxy 
becomes opaque.  The supernova remnants cannot orbit because of their large
collision cross-sections.  They collapse into a central bulge with a quasar
at the center.  The magnetic structures are swept in as well.  If there
is a process in all of this that produces submillimeter radiation, that
radiation is the microwave background.  

Since the 
supernova remnants have high abundances, the bulge gas has high abundances 
and must form high abundance stars.  This can happen both in galaxies that 
are today elliptical or spiral.  These initial quasars continue to be powered 
by infall of gas that is blown off intermediate mass stars when the stars 
climb the giant branch.  This gas is low abundance Population II gas.  It 
dilutes the supernova remnant gas.  This gas forms the disk of spiral galaxies 
so that stars in the disk have abundances initially lower than bulge abundances.  
The oldest population of stars in the disk suffers many globular cluster 
collisions, so it is dispersed into a thick disk.
 
     The quasars eventually run out of fuel.  If later the fuel is
replenished, say by galaxy-galaxy collisions, the quasar can re-ignite.
 
     The activity that we have been describing takes place in the first
10$^{9}$ years.  The time scales are set by orbital and collapse times, and
by stellar evolutionary time scales.  It takes, say, one orbital time
to form the bulge and quasar, and a few orbital times for the mass loss
infall to form the disk.
 
    Since the disk is formed from mass-loss material from Population II stars
in the halo, the mass of the disk gives a lower limit to the mass of the
one- to six-solar-mass primordial Population II stars in the halo and to the
number of white dwarfs.  Each star loses its own mass less the mass of a
white dwarf.
 
    Since the central object and bulge are formed from Population II
supernova remnants, the mass of the central object and bulge (less the
equivalent volume of halo stars) give a lower limit to the mass of the, say,
7 solar mass and greater primordial Population II stars in the halo and
to the number of neutron stars.  Each star loses its own mass less the
mass of the neutron star.
 
\vfill
\eject
\cl{ 9. FORMATION OF D AND $^{3}$HE}
\vs
 
   The initial Population II supernovas produce remnants with magnetic
fields that accelerate cosmic rays.  The halo fills with magnetic structures
and cosmic rays until the supernova remnants collapse to the center to produce 
the quasar and bulge.  The cosmic rays that are not dragged along with the
magnetic fields then decay through normal collisional attrition.
 
     The cosmic rays interact with the primordial neutrino background to 
undergo ladder transmutations to higher or lower elements.  In particular 
a small fraction of $^{4}$He cosmic rays are transformed into $^{3}$He
cosmic rays ($^{4}$He + $\bar{\nu}_{e}$ $\to$ $^{4}$H + e$^{+}$ = 
$^{3}$H + n + e$^{+}$ $\to$ $^{3}$He + n + e$^{+}$ + e$^{-}$ + $\bar{\nu}_{e}$,
and similarly for $\bar{\nu}_{\mu}$ and $\bar{\nu}_{\tau}$) 
and perhaps into D ($^{4}$He + $\bar{\nu}_{e}$ $\to$ 
$^{4}$H + e$^{+}$ = $^{2}$H + n + n + e$^{+}$, if possible).  Through 
collisions $^{3}$He cosmic rays spall into D + p.  Thus D and $^{3}$He are
Population II artifacts and their abundances are a measure of Population II
supernova activity.

     Massive, relatively 
abundant even element cosmic rays are partially transmuted to odd element cosmic 
rays ((Z,A) + $\bar{\nu}_{e}$ $\to$ (Z--1,A) + e$^{+}$; (Z,A) + ${\nu}_{e}$ 
$\to$ (Z+1,A) + e$^{-}$).
 
 
\vfill
\eject
\cl{ 10. FORMATION OF VOIDS AND LARGE SCALE STRUCTURE}
\vs
 
      Next we consider radiatively-driven expansion.  Primordial galaxies produce
a tremendous amount of radiation.  Any galaxy that is a spiral now originally
had most of its mass in massive stars.  A 10$^{12}$ \msun\  spiral galaxy
produces, say, 10$^{11}$ supernovas yielding 10$^{62}$ ergs.  The precurser
stars radiate even more during their lifetimes, say 10$^{63}$ ergs.  There
might be 3x10$^{11}$ intermediate mass stars that radiate 10$^{63}$ ergs and
end up as white dwarfs.  In addition the quasar itself produces 10$^{46}$-
10$^{47}$ ergs s$^{-1}$ for say 3x10$^{8}$ years or about 10$^{63}$ ergs.  There is
also a great deal of energy from the collapse that heats the gas and is eventually
radiated away, partly by the quasar.  If half the large galaxies are spirals, it is
easy to produce 10$^{51}$ ergs \msun$^{-1}$\  averaged over all galaxies.
[Neutrinos produced by the supernovas add up to a similar amount of energy.]
 
      During the first billion years galaxies are much closer together
than now.  If that era corresponds to redshifts of say z=10 to z=5, galaxies are
between 11 and 6 times closer than now.  Statistically it is
possible for a large group of galaxies (say 10$^{5}$) to be optically thick
to their own radiation (except for radio).  Any photon emitted at the center
passes through so many spiral galaxies that it must be absorbed, Figure 13.  
Thus the clump of galaxies expands from its own radiation pressure.  Galaxies
with high projected opacity-to-mass ratios, perhaps face-on spirals,
are accelerated the most, followed by all the other spirals.
The elliptical galaxies are dragged along by
gravitational attraction.  A low density region forms and continues to
expand from radiation pressure as long as the galaxies are very bright and until
the clump of galaxies becomes optically thin.  The expansion of the universe
eventually guarantees the latter.  Eventually the role of radiation becomes
insignificant compared to gravity.
 
     Reg\H{o}s and Geller (1991) have shown that some of the small, low-density
expanding regions in a uniform background will continue to expand gravitationally
as the universe expands, Figure 14.  They form voids that collide and merge.  The
collisions produce galaxy clusters, streaming in the void walls, and
eventually the large scale structure that we see today.
 
\vfill
\eject
\cl{ 11. SUMMARY}
\vs
 
A Big Bang universe consisting, before recombination, of a  gas of 
H, D, $^{3}$He, $^{4}$He, $^{6}$Li, and $^{7}$Li ions, electrons, photons,
and massless neutrinos at a density sufficient to
produce a flat universe, will evolve into the universe as we now observe it.
Evolution during the first billion years is controlled by radiation.
 
The universe has evolved as follows since recombination:
 
1) There were pre-existing galaxy-size perturbations.

2) Recombination halved the gas pressure and removed the outward radiative 
acceleration from these perturbations thereby producing an inward impulse.  
The impulse generated waves that interfered and shocked to fill the large 
perturbations with globular-cluster-size perturbations.

3) The smallest perturbations formed superluminous Population III stars whose 
radiation caused
 
4) larger perturbations to implode and form globular clusters of Population II 
stars, and then
 
5) systems of globular clusters suffered radiatively-triggered collapse 
(violent relaxation) into elliptical galaxies, some of which
 
6) evolved to form quasars and spirals that
 
7) gave off so much radiation that, in some places, statistically, voids were
formed by radiation pressure, and then
 
8) void collisions and void walls produced clusters of galaxies and the large
   scale flows and structure that we see today.

9) The microwave background radiation is recent, younger than the galaxies.
 
     The number of Population III stars was very small and they all exploded
so that only remnants are left.  Essentially all matter has been processed
in stars.  The interstellar medium was produced by stars.  The intergalactic 
medium was produced by galaxies.  It is not primordial.
 
All spiral and irregular galaxies that have not been damaged by collisions or 
interactions have large, massive, elliptical halos.
 
     Figure 15 is the table of contents for our galaxy.  Our galaxy has a 
halo containing about 10$^{11}$ neutron stars,
3x10$^{11}$ white dwarfs, visible K and M stars, and 10$^{11}$ slightly evolved
low mass stars (all numbers to astronomical accuracy).  It also has over
10$^{2}$ coeval globular clusters that are the remnants of 10$^{6}$ primordial
globular clusters from which our galaxy was formed.  There is a central,
inactive, quasar surrounded by a bulge of high abundance Population II stars.
Both were made from the first Population II supernova remnants which collapsed
from the halo to the center of the galaxy.  The disk was made from gas lost
by intermediate mass Population II stars in the halo when they evolved up the
giant branch, and that gas subsequently collapsed into the disk and spun up
to conserve angular momentum.  Thus the disk has lower abundances than the
bulge, even though it was formed later.  There were still many globular
clusters at the time of disk formation so many disk stars were scattered by
collisions with globular clusters and formed a thick disk population.  There
are stars in the halo and globular clusters that were formed in the disk or
bulge and were accreted by globular clusters and carried into the halo.  The
stars in globular clusters need not be siblings, coeval, or Population II.
Non-primordial globular clusters could have been formed in the bulge, the
disk, or in collapsing gas clouds.

During the first billion years evolution 
was controlled by changing matter into radiation in massive stars.  Gravity 
became dominant only after these initial bursts of radiation were exhausted.
\vs
\vs
This work was supported in part by NASA grants NAG5-824 and NAGW-1486.
\vs
\vs
\cl{REFERENCES}
\vs
 
\ni Klein, R.I., Sandford, M.T.,II, \& Whitaker, R.W. 1983, ApJ, 271, L69
 
\ni Kurucz, R.L. 1992, Comments on Astrophysics, 16, 1-15.
 
\ni Kurucz, R.L. 1995, ApJ, 452, 102-108.
 
\ni Lubowich, D.A., Pasachoff,J.M., Galloway, R.P., Kurucz, R.L., and Smith, V.V. 

1994, BAAS, 26, 1479.

\ni Reg\H{o}s, E. \& Geller, M.J. 1991, ApJ, 377, 14-28.
 
\ni Sandford, M.T.,II, Whitaker, R.W., \& Klein, R.I. 1982, ApJ, 260, 183-201.

\ni Sandford, M.T.,II, Whitaker, R.W., \& Klein, R.I. 1984, ApJ, 282, 178-190.
 
\ni Sasselov, D. and Goldwirth, D. 1995, ApJL, 444, L5-L8.

\vfill
\eject
\cl{FIGURE CAPTIONS}
\vs
Figure 1.  Big Bang abundances work if the density is chosen to
close the universe.  Observations:  He, Sasselov and Goldwirth (1995); 
D upper limit, Lubowich et al (1994); Li, Kurucz (1995). 

Figure 2.  Schematic globular-cluster-size perturbations superposed
on top of galaxy-size perturbations.  

Figure 3.  Simulations of radiatively-driven implosions of Population I
clouds indicate the plausiblity of forming a globular cluster by surrounding
a cloud with hot stars.  

Figure 4 qualitatively demonstrates that small radiative accelerations
are sufficient to trigger the collapse of a universe full of globular
clusters into a universe full of elliptical galaxies.  I borrowed the program
from Reg\H{o}s that she used to model void formation (Reg\H{o}s and Geller 1991).
The universe is periodically tesselated into cubes with constant density of 
globular clusters, 128**3 per cube.  Each cube is subdivided into 8
parallellopipeds as shown in the upper left.  This is an arbitrary choice
intended not to look like galaxy precursers.  All the surfaces of all the 
parallellopipeds are given a small inward velocity as would be produced
by excess supernovas at the the surfaces.  The initial condition is zero  
gravitational force.  The small motion of the surface globular
clusters is enough to cause violent relaxation into a galaxy, except in one
case where neighboring galaxies cause the smallest object to disintegrate
and then assimilate its remains.

Figure 5.  Schematic evolution of galaxy of 1/2 \msun\  stars.

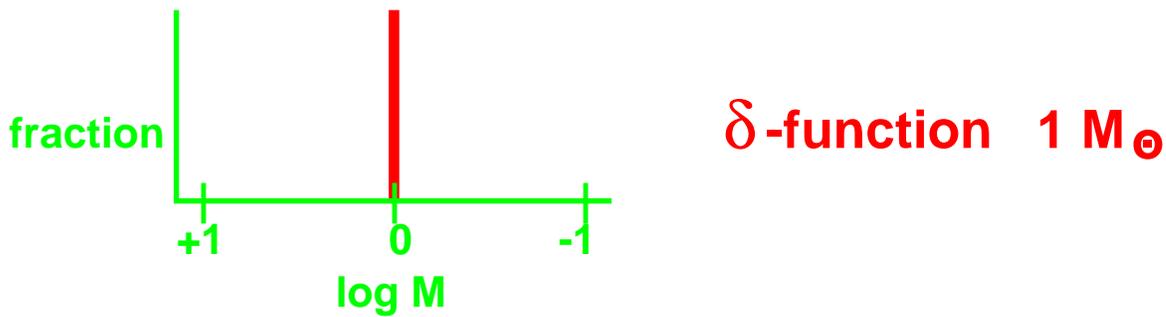
Figure 6.  Schematic evolution of galaxy of 1 \msun\  stars.

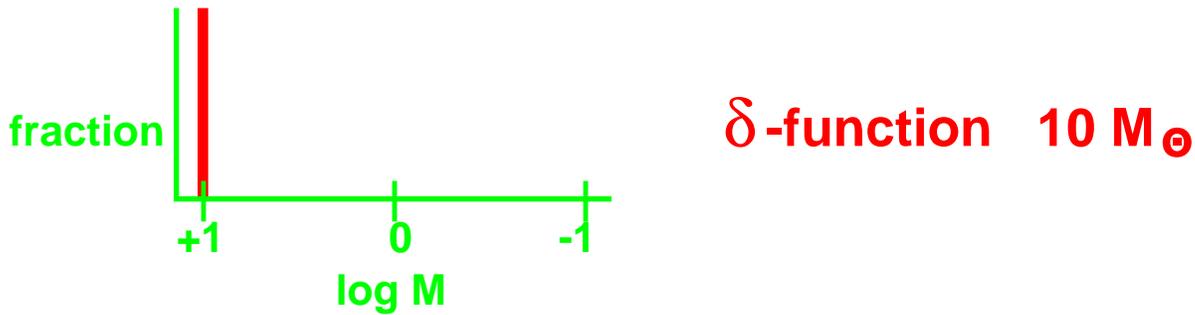
Figure 7.  Schematic evolution of galaxy of 10 \msun\  stars.

Figure 8.  Schematic evolution of galaxy with distribution function peaking 
at 2/3 \msun\  stars.

Figure 9.  Schematic evolution of galaxy with distribution function peaking 
at 1 \msun\  stars.

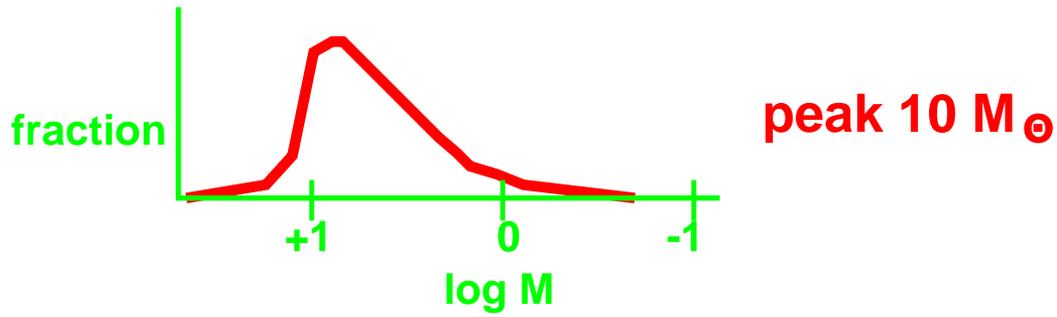
Figure 10.  Schematic evolution of galaxy with distribution function peaking 
at 10 \msun\  stars.

Figure 11.  Evolution of our galaxy.

Figure 12.  Isolated galaxy classification as a function of 
galaxy mass and of stellar mass distribution function peak.

Figure 13. The galaxies are so close together that for some large samples 
any ray out from the center intersects enough spiral galaxies to be absorbed.  
The collection of galaxies is optically thick.

Figure 14. Reg\H{o}s and Geller (1991) showed that starting with a uniform
density universe, one could evolve voids and large scale structure by
removing half the matter from small spheres and redistributing it in
expanding shells.

Figure 15.  Table of contents of our galaxy.

\vfill\bye